\documentstyle[11pt,aasms,tighten,flushrt]{article}
\def \cm{~\rm{cm}}
\def \s{~\rm{s}}
\def \km{~\rm{km}}
\def \g{~\rm{g}}
\def \AU{~\rm{AU}}
\def \yr{~\rm{yr}}
\def \K{~\rm{K}}
\begin{document}
%\normalsize
%\small

\title{
ROTATION, PLANETS, AND THE ``SECOND PARAMETER''  \\
OF THE HORIZONTAL BRANCH }
%\vspace*{2.0cm}

\author{ 
Noam Soker \& Amos Harpaz}
\affil{
Department of Physics, University of Haifa at Oranim\\
%Mathematics-Physics\\
Oranim, Tivon 36006, ISRAEL \\
soker@physics.technion.ac.il \\      
phr89ah@vmsa.technion.ac.il        }

%\clearpage 

\bigskip
\bigskip
\centerline {\bf ABSTRACT}
%\bigskip
%\begin{abstract}

We analyze the angular momentum evolution from the red giant branch
(RGB) to the horizontal branch (HB) and along the HB.  Using rotation
velocities for stars in the globular cluster M13, we find that the
required angular momentum for the fast rotators is up to $1-3$ orders
of magnitude (depending on some assumptions) larger than that of the sun.
Planets of masses up to five times Jupiter's mass and up to an initial
orbital separation of $\sim 2 \AU$ are sufficient to spin-up the RGB
progenitors of most of these fast rotators. Other stars have been spun-up
by brown dwarfs or low-mass main sequence stars. Our results show that
the fast rotating HB stars {\it must} have been spun-up by
planets, brown dwarfs, or low-mass main sequence stars, while they
evolved on the RGB.  We argue that the angular momentum considerations
presented in this paper further supports the ``planet second parameter''
model. In this model, the ``second parameter'' process, which determines
the distribution of stars on the HB, is interaction with low-mass
companions, in most cases gas giant planets, and in a minority of 
cases with brown dwarfs or low-mass main sequence stars.  The masses and
initial orbital separations of the planets (or brown dwarfs or low-mass
main sequence stars) form a rich spectrum of different physical
parameters, which manifests itself in the rich varieties of HB morphologies
observed in the different globular clusters.

%\end{abstract}

%{\it Subject heading:} 

%\clearpage 

% ======================================================================
\section{INTRODUCTION}
% ======================================================================

 In recent years $\sim 20$ Jupiter-like planets have been discovered
around main sequence stars.
 This further increased the great interest in the mechanisms
for planet formation, in particular at close orbital separations.
 Another aspect of the presence of planets is their influence
on the central star and its evolution.
 The processes, some of which were studied before the discovery of
extrasolar planets, include the transfer of a massive planet into a 
star through the accretion from the envelope of a giant star
(Eggleton 1978; Livio 1982; Livio \& Soker 1984; Soker, Harpaz, \& Livio  
1984), formation of SiO masers in the magnetospheres of several 
gas-giant planets around Mira-stars (Struck-Marcell 1988),
formation of R Coronae Borealis stars through enhanced mass loss rate
caused by a planet inside the envelope of an evolved star
(Whitney, Soker, \& Clayton 1991), the formation of elliptical
planetary nebulae by planets spiraling-in inside the envelop of AGB
stars (Soker 1992; 1996; 1997),
evaporation of planets deep in the envelopes of giants
(Harpaz \& Soker 1994; Siess \& Livio 1999a,b),
and enhancing metallicity by planet
evaporation inside main sequence stars (Sandquist {\it et al.} 1998).

 In this paper we examine the evolution of angular momentum of stars
evolving from the RGB to the HB, after being spun-up on the RGB
by planets.
To account for fast rotating HB stars Peterson, Tarbell \& Carney (1983)
already mentioned the possibility that planets can spin-up
RGB stars, though later they abandoned this idea.
  Following the reasons given by Soker (1998), we think that planets
are required to account for the fast rotating HB stars.
 Pinsonneault, Deliyannis, \& Demarque (1991) argue that single
stars can account for the fast HB rotators.
 We disagree with them on two points.
  First, they require that the cores of main sequence stars rotate much
faster than the envelopes.
 Soker (1998; also Behr {\it et al.} 1999) brings causes to question
this assumption.
 Second, they require the core of the RGB star to possess higher
angular momentum than the envelope.
 Since the core of RGB stars is very small this seems to us an
unreasonable assumption.
 Possible support for our claim is the rotation velocity plot of the
globular cluster M13 presented by Peterson, Rood, \& Crocker (1995)
and Behr {\it et al.} (1999), which show that hotter HB
($10,500 \lesssim T_{\rm eff} \lesssim 20,000 \K$)
stars have lower rotation velocities than cooler
($8,000 \lesssim T_{\rm eff} \lesssim 10,000 \K$) stars.
 Not only do the hotter HB stars in these sample rotate slower, they
are also smaller and less massive than the cooler stars,
hence they have much less angular momentum than the cooler HB stars.
 If the angular momentum comes from the core after most of the envelope
has been lost on the RGB, we would expect to find the same
angular momentum distribution along the HB.
 This problem with a single star evolution was noted by Behr
 {\it et al.} (1999) as well. 
 We attribute the much lower angular momentum of hotter HB stars to
angular momentum loss on the RGB, and possibly during the contraction
toward the HB and on the HB.
  In our proposed scenario, stars which do not interact with any massive
planet, brown dwarf, or a low-mass main sequence star rotate very slowly
and do not lose much mass on the RGB.
 Such are, for example, the slowly rotating RR Lyrae variables
(Peterson, Carney, \& Latham 1996).

 We also accept the scenario proposed by Soker (1998) to account for
some anomalies on the HB, and the view that the presence of planets
and brown dwarfs is one of the factors which determine the
``second parameter'' (see also Siess \& Livio 1999b).
  The HB morphologies, i.e.,
the distribution of stars on the HB of a stellar system,
differ substantially from one globular cluster to another. 
 It has long been known that metallicity is the main factor which
determines the location of HB stars on the HR diagram. 
 For more than 30 years, though, it has been clear that another factor is
required to explain the variation in HB morphologies among 
globular clusters with similar metallicity 
(Sandage \& Wildey 1967; van den Bergh 1967; see reviews by 
Fusi Pecci \& Bellazzini 1997; Rood, Whitney, \& D'Cruz 1997;
Rood  1998; de Boer 1998).
 This factor is termed the {\it second parameter} of the HB.
 It seems that stellar companions alone cannot be the second parameter
(e.g., Rich {\it et al.} 1997), nor any other single factor which has
been examined (e.g., Ferraro {\it et al.} 1997 and references therein). 
 We think that the presence of low mass stars and of {\it planets}
(or brown dwarfs) is the main second parameter factor (but probably not
the only one), with planets occurring more frequently.

 The main role, but not the only one, of the planets is the spinning-up
of the RGB envelope.
 It is commonly accepted now that rotation has a connection with the second
parameter, probably through its role in determining the mass loss
on the RGB, directly or indirectly.
 We agree with this assertion, and further claim that the source of
angular momentum in many cases is the interaction with a planet.
  The arguments to support this claim, the scenario proposed by
Soker (1998), and the aim of the present work are summarized in $\S 2$.
 In $\S 3$ we follow the evolution of angular momentum
from the RGB to the HB and along the HB, and apply the results to the
stars of the globular cluster M13. 
 Our summary is in $\S 4$.

% ======================================================================
\section{PLANETS, ROTATION, AND MASS LOSS}
% ======================================================================

 In recent years it has become a common view that the second parameter
determines the HB morphology by regulating the mass loss on the
RGB  (e.g., Dorman, Rood, \& O'Connell 1993; D'Cruz {\it et al.} 1996;
Whitney {\it et al.} 1998; D'Cruz {\it et al.} 2000; Catelan 2000). 
  According to this view, the extreme HB (EHB) stars for example, lose 
almost all their envelope  while still on the RGB
(Dorman {\it et al.} 1993; D'Cruz {\it et al.} 1996). 
 On the RGB, which is the stage prior to the HB and before core helium 
ignition, the star is large and luminous and has a high mass loss rate.
  Sweigart \& Catelan (1998; also Sweigart 2000 and
Moehler, Sweigart, \& Catelan 1999) claim that mass loss on the RGB by
itself cannot be the second parameter, and it should be supplied by
another process, e.g., rotation, or helium mixing, which requires
rotation as well. 
 They term the addition of such a process a ``noncanonical scenario''.
Behr {\it et al.} (1999) find the second parameter problem to be
connected with rotation, and note that single star evolution cannot
explain the observed rotation of HB stars, even when fast core
rotations are considered.
  The rich variety of HB morphologies (e.g.,  Catelan {\it et al.} 1998)
suggests that there is a rich spectrum in the factor(s) behind the
second parameter.

 We argue, following Soker (1998), that these factors are the
masses and orbital separations of the companions, which in most cases
are planets and brown dwarfs, and in the remaining cases they are
low mass main sequence stars.
 That is, we claim, like Siess \& Livio (1999b), that the
``noncanonical scenarios'' (e.g., Sweigart \& Catelan 1998; Sweigart 2000),
involves interaction with planets.
 There is no reason to reject the idea that planets similar to those
discovered recently in the solar neighborhood exist in old star clusters
(Ford {\it et al.} 1999).
 Ford {\it et al.} (1999) argue for the presence of a primordial
star-planet system in the PSR B1620-26 system in the globular cluster M4,
holding that this finding may hint at the presence of many planets in
globular cluster.
  The companions can influence the mass loss rate in several ways
(Soker 1998; Siess \& Livio 1999b).
First, as the companion spirals-in inside the RGB stellar envelope it
deposits (positive) gravitational energy which can remove part of the
envelope.
 Second, the companion spins-up the envelope. Angular momentum may
enhance mass loss rate, mainly near the equator, by, e.g., enhancing
magnetic activity.
 Third, the rotation may mix helium to the envelope
(Sweigart 1997a,b; Sweigart \& Catelan 1998).
 The higher helium abundance increases the RGB tip luminosity,
hence total mass loss on the RGB, leading to the formation of blue HB stars. 
 Sweigart (1997a,b) suggests that this can explain the second parameter, 
though he does not mention the required angular velocity and how his model
accounts for the different groups on the HB.   
  Whitney {\it et al.} (1998) show that the mixing mechanism of Sweigart
cannot explain the data presented in their paper, while  
Ferraro {\it et al.} (1997), Grundahl {\it et al.} (2000),
and Charbonnel, Denissenkov, \& Weiss (2000) discuss other
difficulties with the mixing mechanism proposed by Sweigart being always
important.
 In the scenario proposed by Soker (1998), helium mixing is only one of
several processes caused by planets, so the comments
raised by these works do not contradict the model. 

 In the scenario proposed by Soker (1998), the planets which enter the
envelope of RGB stars can end in one of three ways (Livio \& Soker 1984;
also Seiss \& Livio 1999b):
($i$) evaporation of the planet in the envelope, before the entire envelope
is lost;
($ii$) collision of the planet with the core, i.e., the planet
overflows its Roche lobe, before the entire envelope is lost  
(a planet of radius $\sim 0.1 R_\odot$ and a mass of 
$\sim 0.01 M_\odot$ will overflow its Roche lobe when at 
$\sim 1 R_\odot$ from the core);   
and ($iii$) expelling the envelope while the planet survives the 
common envelope evolution. 
 Soker suggests that these three routes may explain the three subgroups
found by Sosin {\it et al.} (1997) in the blue HB of the globular
cluster 2808.
 In the original scenario, the three routes were determined mainly by the
secondary mass $M_s$.
 The first route, evaporation, occurs for  $M_s \lesssim 1 M_J$, where
 $M_J$ is Jupiter's mass; the second route, of core collision, occurs for
$1 M_j  \lesssim M_s \lesssim 10 M_J$, while the third route, of surviving
companions, occurs for $M_s \gtrsim 10 M_J$. 

  Soker claims also that:
``It is clear that not only planets play a role in the proposed second
parameter mechanism due to companions, since, for example, stellar binary 
companions and occasional collisions with passing stars can also influence 
the mass loss on the RGB''.
 Soker (1998) estimates that in $\sim 40 \%$ of the cases the interaction
of the RGB star is with a stellar companion rather than with a planet.
When publishing the model Soker (1997; 1998) was not aware of
a new class of objects named EC14026
(e.g., Kilkenny {\it et al.} 1997; Koen {\it et al.} 1997;
Koen {\it et al.} 1998, and references therein)
and their relation to EHB stars (Bono \& Cassisi 1999).
  The EC14026 stars are sdB stars which have very rapid light variations,
resulting from pulsation in several modes.
 Low mass binary companions have been detected in several of these stars.
PG 1336-018, for example, has a secondary of mass $\sim 0.15 M_\odot$
with an orbital period of $0.1$ days (Kilkenny {\it et al.} 1998).
 Maxted {\it et al.} (2000) find the orbital periods and minimum
companion masses of two sdB stars: $0.63 M_\odot$ and $8.33$ days for
PG 0940+068, and $0.09M_\odot$ and $0.599$ days for PG 1247+554. 
 For others, the companion, if it exists, is limited to spectral type
M0 or later (e.g., PG 1605+072, Koen {\it et al.} 1998;
PG 1047+003, O'Donoghue {\it et al.} 1998).
 For these systems, we suggest that the companion may be a brown dwarf
or a massive planet as well.
 Allard {\it et al.} (1994) estimate that $\sim 60 \%$
of hot B subdwarfs have binary stellar companions.
 Here again, the stars with no stellar companion may have a
substellar companion.
 Green {\it et al.} (1998) argue that their ``investigations in open
clusters and the field strongly suggest that most
metal-rich BHB [blue HB] stars are in binaries''.
  The EC14026 members with surviving low mass stellar companions support
the scenario proposed by Soker (1997, 1998), and suggest that this scenario 
can be extended to include low-mass main sequence stars.
Therefore, the third route, where the companions survive at small
orbital separations, should include low-mass stars, as well as massive
planets and brown dwarfs.
 When low mass stellar companions are included, the lower limit on the mass
of surviving secondaries can be raised, without causing difficulties to the
model, from $\sim 10 M_J$ to several$\times 10 M_J$. 
 This enriches the varieties of HB stars that can be formed from
binary interactions.

 The main effects of a binary companion as it spirals-in inside the
envelope, or at an earlier phase outside the envelope through tidal 
interaction, are the deposition of gravitational energy and angular 
momentum (Soker 1998). 
 The RGB star, now rotating faster, is likely to increase its mass loss
rate, lose more mass on the RGB, and hence turn into a bluer HB star.
But many blue HB stars retain some of the envelope mass, which means
that the RGB star does not lose its entire envelope. 
 In this paper we examine the evolution of the angular momentum
in these stars, which possess high angular momentum, as they
contract and evolve toward and on the HB.

 Before turning to the calculation of angular momentum evolution we estimate
the angular momentum deposited by a low-mass companion (Soker 1998).
 When the star evolves along the RGB it expands slowly.
 When its radius $R$ becomes $\sim 20 \%$ of the orbital separation $a_0$, 
tidal forces will cause the substellar companion orbit to decay in a time 
shorter than the evolutionary time (Soker 1996), thus forming a common 
envelope phase. 
 The angular velocity of the envelope $\omega$ can be estimated as follows. 
The envelope's moment of inertia is 
$I_{\rm {env}}= \alpha M_{\rm {env}} R^2$,
where $M_{\rm {env}}$ is the envelope's mass and $R$ is the stellar radius
(we neglect the core's moment of inertia since it is very small),
and $\alpha \simeq 0.1$ for upper RGB stars (see next section).
 The final envelope angular momentum $I \omega$ is equal to the planet's  
initial orbital angular momentum 
$J_p=M_p (G M_1 a_0)^{1/2} = M_p \omega_{\rm {Kep}} (a_0R^3)^{1/2}$,
where $\omega_{\rm {Kep}}$ is the Keplerian angular velocity on the 
RGB star's surface, $M_p$ is the planet's mass and $M_1$ is the
primary's total mass.
 The envelope angular velocity is
\begin{eqnarray}
\frac{\omega}{\omega_{\rm {Kep}}}  
= 0.1
\left( \frac{M_p}{0.01 M_{\rm env}} \right)  
\left( \frac{a_0}{R} \right) ^{1/2} 
\left( \frac{\alpha}{0.1} \right) ^{-1} .
\end{eqnarray}
 Wide stellar companions  ($a_0 \lesssim 5 \AU$) can
deposit angular momentum via tidal interaction, leading to similar
effects as those of planets.

% ====================================================================
\section{ANGULAR MOMENTUM EVOLUTION}
% ====================================================================
% ====================================================================
\subsection{Angular Momentum Loss}
% ====================================================================

 Not considering magnetic influence beyond the stellar surface,
and assuming a solid body rotation through the stellar envelope,
the angular momentum loss rate from stars is
\begin{eqnarray}
\dot J_{\rm wind} = \beta \omega R^2 \dot M,
\end{eqnarray}
where $\omega, J, R, M$ are the stellar angular velocity,
angular momentum, radius, and mass, respectively,
and $\beta$ depends on the mass loss geometry.
 For a constant mass loss rate per unit area on the surface $\beta =2/3$,
while for an equatorial mass loss $\beta=1$. 
 The angular momentum of the star is $J_{\rm env}=I \omega$, where $I$
is the moment of inertia given by
\begin{eqnarray}
I = \alpha M_{\rm env} R^2,
\end{eqnarray}
where $M_{\rm env}$ is the envelope mass, and 
we neglect the core's moment of inertia relative to that of the envelop
and the change in the core mass at late RGB stages.
 The change of envelope's angular momentum with time is given by
\begin{eqnarray}
\dot J_{\rm env} = \dot I \omega + I \dot \omega =
\frac {d I }{d M_{\rm env}} \dot M \omega + I \dot \omega.
\end{eqnarray}
Dividing equation (2) by equation (3) multiplied by $\omega$, gives also
\begin{eqnarray}
\frac {d \ln J_{\rm env}}{d \ln M_{\rm env}} = 
\frac {d \ln \omega}{d \ln M_{\rm env}}
+ \frac {d \ln I}{d \ln M_{\rm env}}  
=
\frac {\beta}{\alpha (M_{\rm env})} \equiv \delta .
\end{eqnarray}

 If the structure of the atmosphere does not change much with
mass loss, and the density is given by $\rho \propto r^{-2}$,
then $\alpha = 2/9$ and ${d \ln I}/{d \ln M_{\rm env}} = 1$,
for which we find ${d \ln \omega}/{d \ln M_{\rm env}} = 2$,
for $\beta = 2/3$ (Harpaz \& Soker 1994).
 This case is appropriate for the upper AGB.
 On the RGB, if the envelope mass is not too low, we can take
$\alpha \simeq 0.1$ (see below), for which we find from equation (5) that
the total envelope angular momentum decreases as
(neglecting changes in the core mass)
$J_{\rm env} \propto M_{\rm env}^\delta$, where
$\delta \equiv \beta/\alpha \simeq 6-7$.

% ====================================================================
\subsection{The Angular Momentum of Stars in M13}
% ====================================================================

 We now  analyze the rotation velocities of HB stars in the globular cluster
M13, as presented by Behr {\it et al.} (1999), with the goal of finding
the total angular momentum each star had on the RGB.
 For the purpose of this simple analysis, we can make the following
simplifying assumptions:

(1) We assume that the angular momentum was deposited to the
RGB star before it had lost much of its envelope (or if no interaction
with a stellar or substellar companion occurred, then the angular momentum
is that which the star had on the main sequence).
 We take the envelope mass prior to the intensive mass loss episode
to be $M_0=0.3 M_\odot$, as appropriate for a well developed core on the
upper RGB.
 For the situation where the deposition of angular momentum occurred
after the envelope mass was already reduced to
$M_0=0.2 M_\odot$, for example, the angular momentum in each case
will be reduced by $(2/3)^{\delta}=0.2$ (0.07), for
$\delta = 4$ (6.5).

(2) We neglect core evolution during the intense mass loss rate.
This is reasonable, as most of the mass is lost on the upper RGB.

(3) We assume that the angular momentum evolves according to
equation (5), with a constant value of $\alpha$, and with $\beta=2/3$.

(4) For the envelope mass and luminosity of the HB stars in the
interesting region $7,000 < T_{\rm eff} < 20,000 \K$, we approximate the
models of Dorman {\it et al.} (1993) and D'Cruz {\it et al.} (1996)
for [Fe/H]=-1.5, by taking for the envelope mass (in $M_\odot$)
\begin{eqnarray}
M_{\rm HBe} =1.37 -0.31 \log T_{\rm eff},
\end{eqnarray}
and for the luminosity (in $L_\odot$)
\begin{eqnarray}
\log L =5.14 -0.9 \log T_{\rm eff},
\end{eqnarray}
where $T_{\rm eff}$ is the effective temperature (in K).
The stellar radius is calculated from $L$ and $T_{\rm eff}$.

(5) Like Pinsonneault {\it et al.} (1991) we neglect structural changes
along the considered segment of the HB, and take the moment
of inertia of all HB models to be $I_{\rm HB}=0.01 M_{\rm HBe} R^2$,
i.e., $\alpha_{\rm HB} =0.01$.
  Any relative change in the value of $\alpha_{\rm HB}$ is much smaller
than the uncertainties in the values of $\alpha$ on the RGB, hence of
$\delta$.

 These assumptions allow us to find the present angular momentum of
 the HB stars,
\begin{eqnarray}
J_{\rm HB} = 0.01 M_{{\rm HBe}} R v_r,
\end{eqnarray}
where $v_r$ is the rotation velocity.
The rotation velocity is taken to be the value of $v \sin i$, as given for
M13 by Behr {\it et al.} (13 stars) and Peterson {\it et al.} (1995)
(25 stars; for the 4 stars common to these two works we take
values from Behr {\it et al.}).
 From equation (5), the total angular momentum that the RGB progenitor
had before the intensive mass loss started is, with assumption (1) above, 
\begin{eqnarray}
J_0 = J_{\rm HB} (M_{\rm HBe}/M_0)^{-\delta}.
\end{eqnarray}
 To find the value of $\alpha$ on the RGB we evolved an RGB model.
 The model was based on a previous work by one of us
(Harpaz \& Shaviv  1992), hence we used a solar composition.
 However, since we are interested in the value of $\alpha$, which is the
ratio of the envelope moment of inertia to $M_{\rm env} R^2$, the exact
values of the RGB radius, luminosity, and core mass are not so important.
 Indeed, the value of $\alpha$ changes only little for different radii
in the range $90 - 120 R_\odot$, core masses of $> 0.4 M_\odot$,
and envelope masses in the range $0.05 - 0.3 M_\odot$,
and its value on the upper RGB is $\alpha \simeq 0.1$.
 More accurately, we can approximate the value of $\alpha$ in the
range of $0.03 \lesssim M_{\rm env} \lesssim 0.3 M_\odot$ by 
$\alpha = 0.1 +1.1 \times 10^{-4} (M_{\rm env}/M_\odot)^{-1.6}$.
 As the envelope mass decreases below $\sim 0.03 M_\odot$, the RGB envelope
shrinks, and $\alpha$ increases on a slower rate, becoming
only $\alpha \simeq 0.16$ for $M_{\rm env} = 0.01 M_\odot$.
  We find that an appropriate average value is  $\alpha=0.103$, which gives
$\delta(RGB) = \beta/ \alpha \simeq 6.5$ (for $\beta = 2/3$).
 We use this value for $\delta$ even though it decreases 
(since $\alpha$ increases)
for very low envelope masses, e.g., $\delta \simeq 5.9$ for
$M_{\rm env}= 0.05 M_\odot$, since
for the blue HB stars in the sample used here the required angular momentum
is very large, and the evolution is more complicated.
 That is, other uncertainties in the interaction process of the more massive
companion with the RGB envelope dominate, as we discuss later.
 For example, angular momentum loss may be even larger, since $\beta > 2/3$,
at early stage for the RGB progenitors of the blue HB stars in the sample.

 In Figure 1 we plot the RGB angular momentum for $M_0=0.3 M_\odot$,
and for $\delta =4$ (Fig. 1a; upper panel) and $\delta = 6.5$
(Fig. 1b; lower panel).
 We present the results for $\delta=4$ as well as the expected
 value of $\delta \sim 6.5$, since we find a uniform 
(more or less) distribution in $J_0$ (see Fig. 1a).
 The large symbol in the right-hand side of Figure 1b
stands for a star which had $J_0 = 6.05 \times 10^{51} \g \cm^2 \s^{-1}$,
hence it was reduced by a factor of 10 to fit into the graph.
 The value of $J_0$ will be reduced by a factor of 10 if we take
$M_0=0.17 M_\odot$ and $0.21 M_\odot$, for $\delta=4$ and
$\delta=6.5$, respectively. 
 Of course, the mass of the envelope when angular momentum was deposited
to it changes from one star to another, but we have no way
of telling the value of $M_0$ from the properties of the star on the HB.

 Although the angular momentum on the HB is
$J_{\rm HB} \lesssim J_\odot$, where $J_\odot$ is the angular momentum
of the sun, as noted already by Pinsonneault {\it et al.} (1991), the
angular momentum required on the RGB for our preferred scenario
($\delta=6.5$ and $M_0=0.3 M_\odot$)
is up to three orders of magnitude higher.
 Even if we take $M_0=0.2 M_\odot$ and $\delta =4$, the required
angular momentum is up to $\sim 5 J_\odot$.
  As noted by Peterson {\it et al.} (1983), the angular momentum
requires a planet to spin-up the RGB envelopes.
 The orbital angular momentum of a planet (or any other companion of
mass $M_p$)  is
\begin{eqnarray}
J_p =
8 \times 10^{49}
\left( \frac{M_p}{M_J} \right)  
\left( \frac{M_{10}}{0.9 M_\odot} \right)  
\left( \frac{a}{1 \AU} \right) ^{1/2} \g \cm^2 \s^{-1},
\end{eqnarray}
where $M_J$ is Jupiter's mass, $M_{10}$ is the initial stellar mass, and $a$
is the initial orbital separation.
 Jupiter like planets or lighter planets at orbital separations of
$\sim 1 \AU$ can account for the angular momentum of $\sim 30 \%$
of the stars in Figure 1b.
 Allowing planet masses up to $5 M_J$ and orbital separation up to
$\sim 2 \AU$, 34 out of the 38 stars in the sample can be accounted for. 

 The most extreme case, with $J_0=6 \times 10^{51} \g \cm^2 \s^{-1}$,
requires a companion of mass $0.07 M_\odot$ ($0.22 M_\odot$)
at an orbital separation of $\sim 1 \AU$  ($\sim 0.1 \AU$), by our
calculations.
 However, in our calculations we assumed that the deposition of angular 
momentum occurs within a short period, after which mass loss is a 
continuous process. 
 The situation is more complicated when the angular momentum is large. 
 This is seen by calculating the angular velocity of the RGB star
after an angular momentum of $J_p$ is deposited to its envelope 
\begin{eqnarray}
\frac{\omega}{\omega_{\rm {Kep}}}  
= 0.06
\left( \frac{M_1}{0.8 M_\odot} \right)  ^{-1/2}
\left( \frac{M_{\rm env}}{0.3 M_\odot} \right)  ^{-1}
\left( \frac{R}{100 R_\odot} \right) ^{1/2} 
\left( \frac{\alpha}{0.1} \right) ^{-1} 
\left( \frac{J_p}{10^{50} \g \cm^2 \s^{-1} } \right) ^{-1} .
\end{eqnarray}
 Several things should be noted.
First, when $J_p \gtrsim 3 \times 10^{50} \g \cm^2 \cm^{-1}$ the 
angular velocity is  $\omega \gtrsim 0.2 \omega _{\rm Kep}$, and mass 
loss is expected to be concentrated toward the equatorial plane. 
This means that $\beta > 2/3$ (see eq. 2), and $\delta > 6.5$ at
early stages. 
 Second, for a required angular momentum of 
$J_0 \gtrsim 10 \times 10^{51} \g \cm^2 \cm^{-1}$, as for two stars
in Figure 1b,  the companion may bring the envelope to synchronization with
the orbital motion before it enters the envelope. 
This means that the wind may carry even more angular momentum,
since the companion keeps the envelope at a more or less constant
angular velocity even after substantial mass has been lost. 
  Third, if the companion keeps the envelope rotation 
synchronized with the orbital motion until the star goes through a 
helium flash and starts its contraction to the HB,
the star will reach the HB with very high angular velocity
(because of the contraction).  
 This must result in a substantial mass loss during the contraction phase,
which will slow down the star by orders of magnitude
(see next subsection). 
 These three points show that the evolution of precursors of
rotating blue HB stars maybe very complicated. 
 In any case, we conclude that low-mass main sequence stars, or even 
brown dwarfs and planets can explain the angular momentum of all 
stars in this sample.
 Models based on single stars cannot account for the fast rotators on the
HB.

 The main factor that determines the mass loss during a common envelope
evolution is the mass of the companion, while the angular momentum
is determined by the orbital separation as well.
 Assuming that the distributions of initial orbital separations and
initial masses of the companions are independent, we expect that the
most massive companions will cause higher mass loss rate as well
as deposit more angular momentum.
 This is compatible with the distribution on Figure 1b, where
the RGB progenitor of  HB stars which have less envelope mass had
more angular momentum.

% ====================================================================
\subsection{Angular Velocity Evolution along the HB}
% ====================================================================

 An HB star slows down very fast with mass loss on the HB, since
$\alpha \lesssim 0.01$ on the HB.
For example, for $\alpha=0.01$ the star will slow down by a factor
of 2 (30) by losing only $1 \%$ ($5 \%$) of its envelope mass.
 Therefore, envelope rotation on the HB, even if enhancing the mass
loss rate, will not change the evolution on the HB much since the
 star will slow down before losing much mass.
 As an example let us examine the rotation velocities of M13
(Fig. 7b of Peterson {\it et al.} 1995).
 Hot stars ($10,500 \lesssim T_{\rm eff} \lesssim 13,000 \K$) have rotation
velocities of $\sim 10 \km \s^{-1}$.
 They cannot be the descendants of the cooler
($8,000 \lesssim T_{\rm eff} \lesssim 10,000 \K$) fast rotating,
$\sim 40 \km \s^{-1}$, stars.
 To move by $\sim 2000 \K$ to the left on the HR diagram (becoming hotter)
on this part of the HB requires the envelope to lose $\sim 20 \%$ of its
mass [e.g., D'Cruz {\it et al.} 1996; eq. (6) above].
 For $\alpha \simeq 0.01$ on this part of the HB 
(Pinsonneault {\it et al.} 1991), such a mass loss will reduce the 
angular velocity almost to zero.  
 
 There are no observations to indicate the mass loss rate of HB stars.
 The Reimers (1975) formula is not a good estimate for the
mass loss of HB stars (Koopmann {\it et al.} 1994).
 Instead, Koopmann {\it et  al.}, as well as Demarque \& Eder (1985)
take a mass loss rate up to $10^{-9} M_\odot \yr^{-1}$.
 A higher upper mass loss rate of $3 \times 10^{-9} M_\odot \yr^{-1}$
was assumed for HB stars by Yong \& Demarque (1997).
 Following these uncertainties, we stay with the Reimers (1975) mass loss
rate 
\begin{eqnarray}
\dot M \simeq 2 \times 10^{-11}  \eta
\left( \frac{L}{30 L_\odot}\right)
\left( \frac{R}{R_\odot} \right)
\left( \frac{M}{0.5 M_\odot} \right)^{-1}
M_\odot \yr^{-1}, 
\end{eqnarray}
where $L, R, M$ are the stellar luminosity, radius, and mass, 
respectively, and $\eta$  is a constant of  order  unity.
 The lifetime on the HB is $\sim 10^8 \yr$.
 According to the above mass loss rate, HB stars will lose 
$\sim 2 \%$ of their envelope mass during their lifetime on the
HB.
This will not change the location on the HB much, but by taking
$\alpha = 0.01$ in equation (5), we find that the mass loss will slow
down the stellar rotation by a factor of $\sim 4$. 
  
% ====================================================================
\section{SUMMARY}
% ====================================================================

 Rotation along the RGB may increase the total mass loss,
increase the core mass at helium flash (Mengel \& Gross 1976),
and mix helium to the envelope (Sweigart \& Catelan 1998, and references
therein).
 Mengel \& Gross (1976) find that the rotation velocity of
the core should be $\omega_{\rm core} > 2 \times 10^{-4} \s^{-1}$, i.e.,
two orders of magnitude faster than the solar rotation velocity,
in order to substantially influence the evolution along the RGB.
 We think that such high rotation velocities, even of the core alone,
can be attained {\it only} via the interaction with a gas giant planet,
a brown dwarf, or a low-mass main sequence star.
 This led us, following Soker (1998; also Siess \& Livio 1999b),
to propose that planets are the main ``second parameter'' factor of the
HB, although planets alone cannot explain all blue HB stars.
 Extreme HB stars seem to require the interaction with a low-mass stellar
companion, through a common envelope evolution or via tidal interaction
with the companion outside the envelope.
 That stellar companions can lead to the formation of sdB stars
was suggested by Mengel, Norris, \& Gross (1976), through mass transfer.
 We here invoke different processes, namely tidal spin-up or
common envelope evolution.
 Support for the common envelope evolution is the sdB star
PG 1336-018 which has a secondary of mass $\sim 0.15 M_\odot$
with an orbital period of $0.1$ days (Kilkenny {\it et al.} 1998;
see also  Maxted {\it et al.} [2000] for the sdB binary system
PG 1247+554).
 
In this paper we analyze the angular momentum evolution from the
RGB ro the HB and along the HB.
 Using rotation velocities for stars in the globular cluster M13,
we find that the required angular momentum for the fast rotators is
up to $1-3$ orders of magnitude (depending on some assumptions) larger
than that of the sun.
 On the other hand, planets of masses up to five times Jupiter's mass
and up to an initial orbital separation of $\sim 2 \AU$
are sufficient to spin-up the RGB progenitors of most of these
fast rotators (other had been spun-up by brown dwarfs or
low-mass main sequence stars).
 Our results show that the fast rotating HB stars {\it must} have been
spun-up by companions, planets, brown dwarfs, and low-mass main sequence
stars, while they evolve on the RGB.
 
 We argue that the angular momentum considerations presented in this
paper further support the ``planet second parameter'' model.
In this model, the ``second parameter'' process of the HB is
interaction with low-mass companions, in most cases gas giant planets,
and in a minority of the cases with brown dwarfs or low-mass main sequence
stars.
  The masses and initial orbital separations of the planets (or
brown dwarfs or low-mass main sequence stars) form a rich spectrum
of different physical parameters, which manifests itself in the rich
varieties of HB morphologies observed in the different globular
clusters and elliptical galaxies.

% ====================================================================
\acknowledgments
This research was supported in part by grants from the 
Israel Science Foundation and the US-Israel Binational Science Foundation.

{\bf FIGURE CAPTIONS}

\noindent {\bf Figure 1:}
 The total angular momentum of the progenitor RGB star prior to
the intensive mass loss vs. effective temperature of the descendant HB
star.
 The total angular momentum for each RGB star is calculated by equation
(9) with $M_0=0.3 M_\odot$ and a constant value of $\delta$.
(a) For $\delta=4$ (upper panel).
(b) For $\delta=6.5$ (lower panel).
  The large symbol in the right-hand side stands for a star which had
$J_0 = 6.05 \times 10^{50} \g \cm^2 \s^{-1}$,
hence it was reduced by a factor of 10 to fit into the graph.

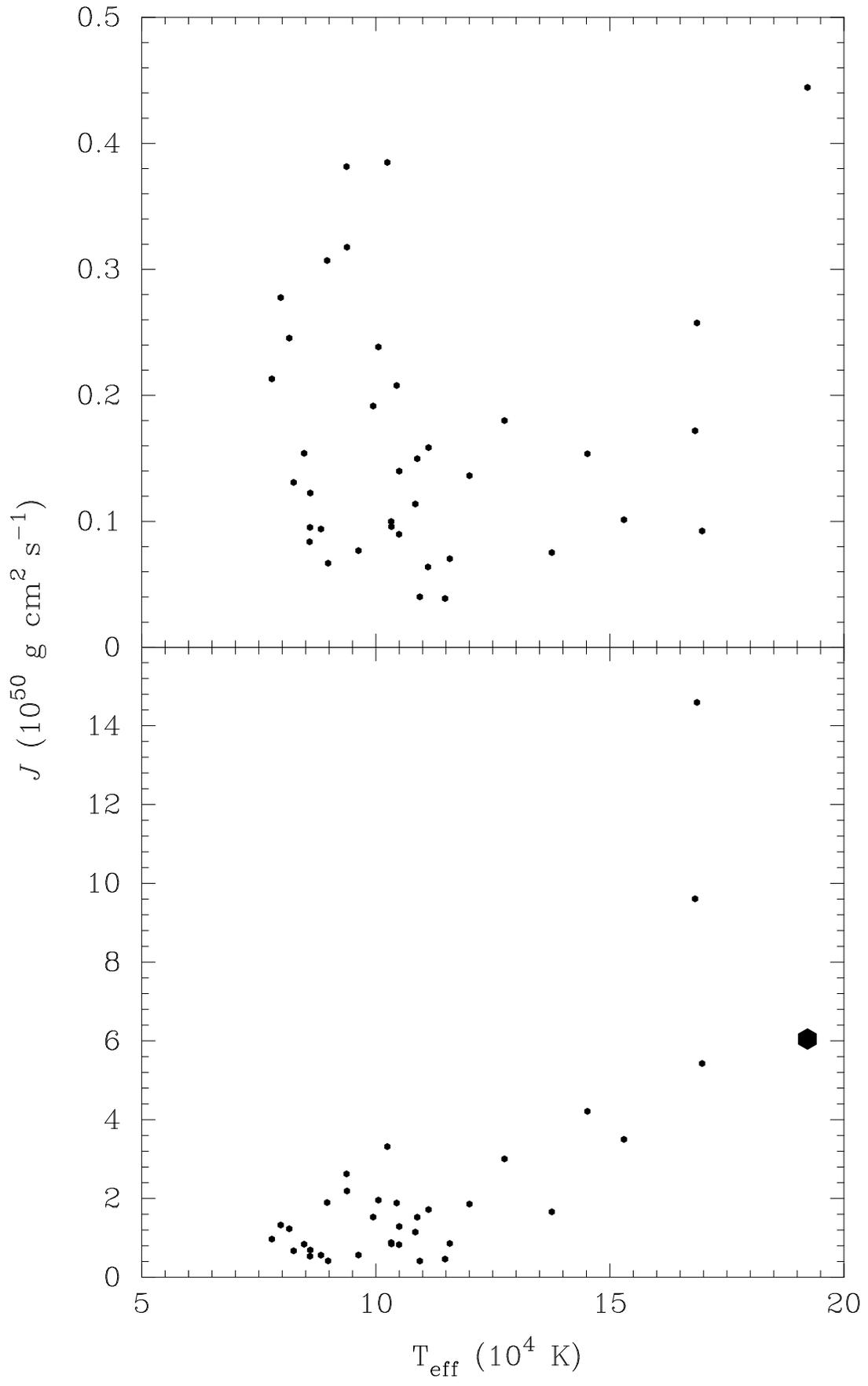
\end{document}